# A Comprehensive *Ab Initio* Study of Electronic, Optical and Cohesive Properties of Silicon Quantum Dots of Various Morphologies and Sizes up to Infinity

**Shanawer Niaz[1,2*] and Aristides D. Zdetsis[1,3]**

[1] Molecular Engineering Laboratory, at the Department of Physics, University of Patras, Patras, GR-26500, Greece

[2] Department of Physics, Bilkent University, Ankara 06800, Turkey

[3] Institute of Electronic Structure and Laser, Foundation for Research and Technology Hellas, Vassilika Vouton, P.O. Box 1385, Heraklion, Crete GR 71110, Greece

[*]To whom correspondence should be addressed

**ABSTRACT**

We present a comprehensive and integrated model-independent *ab initio* study of the structural, cohesive, electronic, and optical properties of silicon quantum dots of various morphologies and sizes in the framework of all-electron "static" and time-dependent density functional theory (DFT, TDFT), using the well-tested B3LYP and other properly chosen functional(s). Our raw *ab initio* results for all these properties for hydrogen passivated nanocrystals of various growth models and sizes from 1 to 32 Ångstroms, are subsequently fitted, using power-law dependence with judicially selected exponents, based on dimensional and other plausibility arguments. As a result, we can reproduce with excellent accuracy not only known experimental and well-tested theoretical results in the regions of overlap, but we can also extrapolate successfully all the way to infinity, reproducing the band gap of crystalline silicon with almost chemical accuracy as well as the cohesive energy of the infinite crystal with very good accuracy. Thus, our results could be safely used, among others, as interpolation and extrapolation formulas not only for cohesive energy and band gap, but also for interrelated properties, such as dielectric constant and index of refraction of silicon nanocrystals of various sizes all the way up to infinity

**Keywords:** Silicon Nanocrystals, Quantum Dots, Cohesive properties, Electronic properties, Optical properties, DFT calculations, TDDFT calculations, Surface reconstruction.

# 1. INTRODUCTION

Silicon nanocrystals, or quantum dots (due to their zero dimensionality, compared to infinite Si crystals, 3D, films, 2D, or wires, 1D) have attracted a lot of interest over the last years due to their potential band-gap engineering properties. The main reason is that, contrary to the electronic properties, the optical properties of crystalline silicon are rather poor because of the small (smaller than the lower edge of the visible spectrum) and indirect band gap, resulting in phonon-assisted emission. Thus, the optical properties of silicon quantum dots (QD), which are inherently connected with the electronic properties, as well as with the bonding and cohesive properties, have been a very challenging and promising field of research over last decade for obvious technological and scientific reasons.[1-25] The culmination of the silicon quantum dots research occurred with the observation of visible photoluminescence (PL) in porous silicon and silicon nanocrystals.[1] Hence, most of the work in this field has been devoted to understand and tune the visible photoluminescence of the QDs by adjusting and correlating the optical gap with the size (diameter) of the dots,[2-4] not always without inconsistencies and ambiguities,[4] which are related with the difficulty to exactly determine the QD size and the exact morphology and composition of its surface layer.[4] However, with the advancement of technology in recent years, these problems are not so serious. It is widely accepted by now that the visible luminescence of small and pure (oxygen-free) QD samples with well defined diameters, is mainly due to quantum confinement of the corresponding quantum dots.[4,5,6] Yet, since the dot's properties are sensitive to the preparation conditions and the growth environment, several other alternative mechanisms have been also considered in the past for the detailed description of the variation of the gap with size (number of particles or diameter), and surface conditions of the dots, such as free-exciton collision,[7] and impurity luminescent centre mechanism.[8]

The unique, size and composition, tunable electronic and optical properties of Si quantum dots make them very appealing for a variety of applications and new technologies. Examples include LEDs,[9] solid-state lighting displays,[10] and photovoltaics.[11] Being zero dimensional, quantum dots have a sharper density of states than higher-dimensional structures. Their small size also means that electrons do not have to travel as far as with larger particles, thus electronic devices can operate faster. Examples of possible applications taking advantage of these unique electronic properties include transistors[12] and logic gates,[13] and quantum computing,[14] among many others. The small size of quantum dots allows them to go anywhere in the body, making them suitable for different bio-medical applications[15] like medical imaging[16] and biosensors,[17] etc. At present, fluorescence based biosensors depend on organic dyes with a broad spectral width, which limits their effectiveness to a small number of colours and shorter lifetimes to tag the agents. On the other hand, quantum dots can emit the whole spectrum, are brighter and have little degradation over time, thus being superior to traditional organic dyes used in biomedical applications.

In this study three distinct growth models (morphologies) of silicon quantum dots are studied such as elongated, spherical (grown along [111] direction) and reconstructed dots. Spherical QDs with diameters d smaller than 2 nm (d< 2 nm) have been studied by our group earlier[5] with considerable success. These calculations have served, among others, as "yard sticks", especially in the gap-size dependence, in several experimental and theoretical works. In this work we have considered alternative QD morphologies and have expanded their size up to 3.2 nm. In several cases we have also considered alternative modern functionals for comparison, although we have already tested B3LYP with high level *ab initio* results in the past.[5] As was expected, in the framework of the present investigation, we found that there is no need to resort to other type of functional(s). Concerning new morphologies, we have considered in addition to spherical, elongated and reconstructed QDs. We have already introduced spherical silicon quantum dots (d < 2 nm)[5] whereas, in this work, we include large dots (up to 3.2 nm in diameter). For reconstructed dots, Hongdo et al.[18] reported that step and dimmer reconstruction decrease the gap values and modulate the charge distribution, inducing spatial separation of near-gap levels. The predicted induction in spatial separation of HOMO-LUMO can be used for designing efficient solar cells. In this work, special effort has been placed in examining the quantum confinement concept on large quantum dots in which the gaps have been obtained with very high accuracy. Our results verify the quantum confinement dependence and agree with experimental measurements (wherever is possible) in and outside the size range of our calculations, making it possible to successfully extrapolate nanoscale results in the intermediate region all the way to infinite silicon crystal. On the basis of existing (empirical) relationships between gap and dielectric constant or index of refraction[26,27] if one wishes to rely on such methods, one can also obtain estimates of such quantities for a given QD size and morphology.

## 2. TECHNICAL DETAILS

All DFT, TDDFT and frequency calculations were performed with the TURBOMOLE[28] suite of programs for medium size of dots (d < 20 Å). For larger dots of diameter d > 20 Å, calculations were performed in GAUSSIAN 03[29] package because of the number of basis function limitations in TURBOMOLE. All *ab initio* calculations are based on the DFT/B3LYP method, employing the hybrid nonlocal exchange-correlation B3LYP[30] functional. This functional has been shown to efficiently reproduce the band structure of a wide variety of materials, including c-Si, with no need for further numerical adjustments.

The SVP[31] basis set was used for geometry optimization of the larger dots (for computational economy), after which single point calculations of the energy were performed with the TZVP[32] basis (and in some cases def2-TZVPP[33]). In addition to this we have also tested (in selected cases) for possible basis set superposition error by using the counterpoise method. Convergence criteria for the SCF energies and for the electron density (rms of the density matrix), were placed at $10^{-7}$ au, whereas for the Cartesian gradients the convergence criterion was set at $10^{-4}$ au.

## 3. RESULTS AND DISCUSSION

As we mentioned above that we present a very detailed discussion on our results concerning various growth models. Thus we discuss the structural, electronic, cohesive and optical properties respectively using DFT/TDDFT calculations in order to investigate the stability and their size dependence of the silicon quantum dots.

### 3.1. Structural Properties

We construct spherical quantum dots with $T_d$ symmetry by keeping one atom at origin and grow them spherically in [111] direction. The size of the spherical quantum dots, considered here, ranges from 17 to 717 Si atoms with 36 to 300 H atoms (a total of 1017 atoms in largest dot). The diameter of small ($Si_{17}H_{36}$) and large ($Si_{717}H_{300}$) cluster is 9.62 Å and 30.95 Å respectively. The optimized structures of spherical quantum dots are shown in Figure 1.

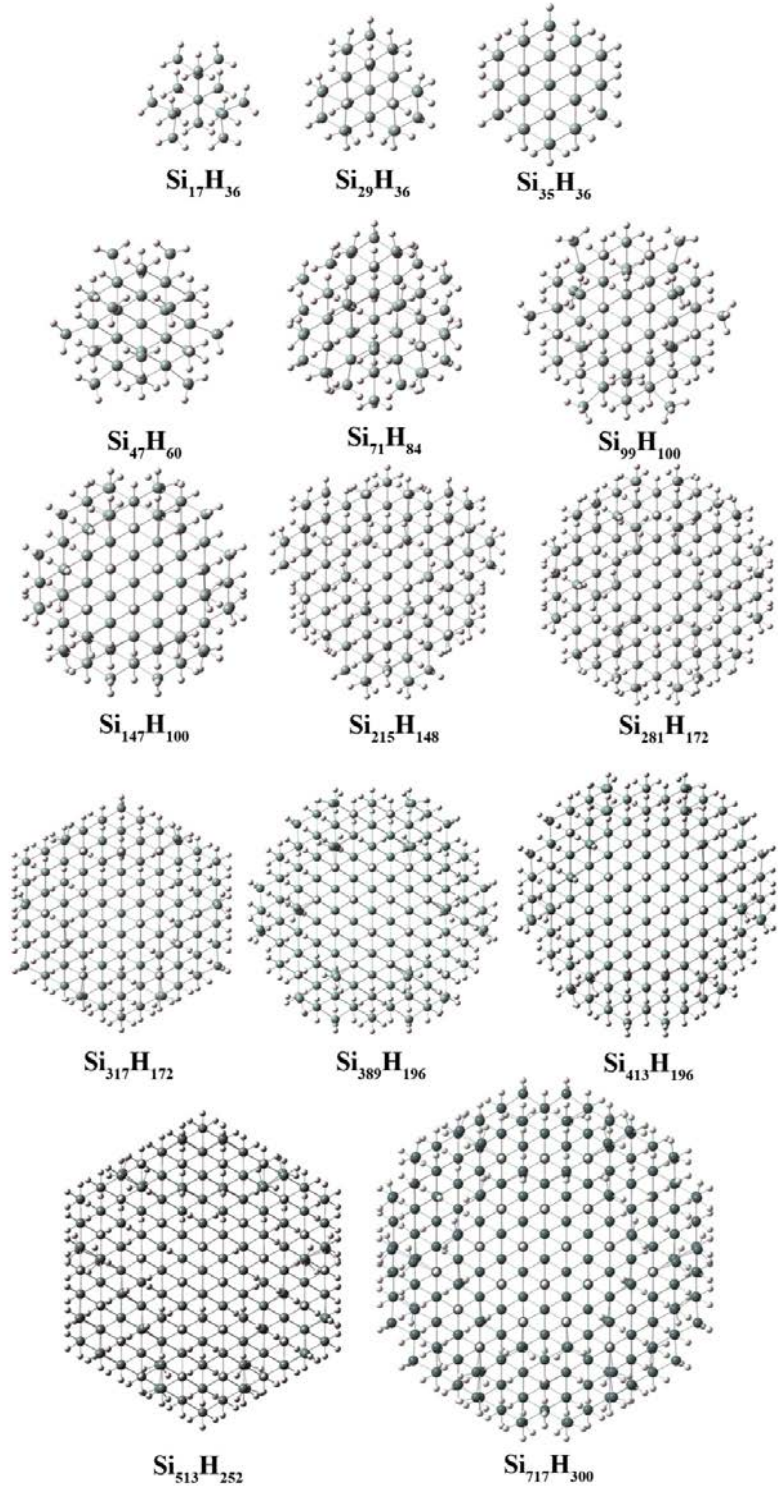

Figure 1: The optimized geometries of Si$_{17}$H$_{36}$, Si$_{29}$H$_{36}$, Si$_{35}$H$_{36}$, Si$_{47}$H$_{60}$, Si$_{71}$H$_{84}$, Si$_{99}$H$_{100}$, Si$_{147}$H$_{100}$, Si$_{215}$H$_{148}$, Si$_{281}$H$_{172}$, Si$_{317}$H$_{172}$, Si$_{389}$H$_{196}$, Si$_{413}$H$_{196}$, Si$_{513}$H$_{252}$ and Si$_{717}$H$_{300}$ spherical quantum dots.

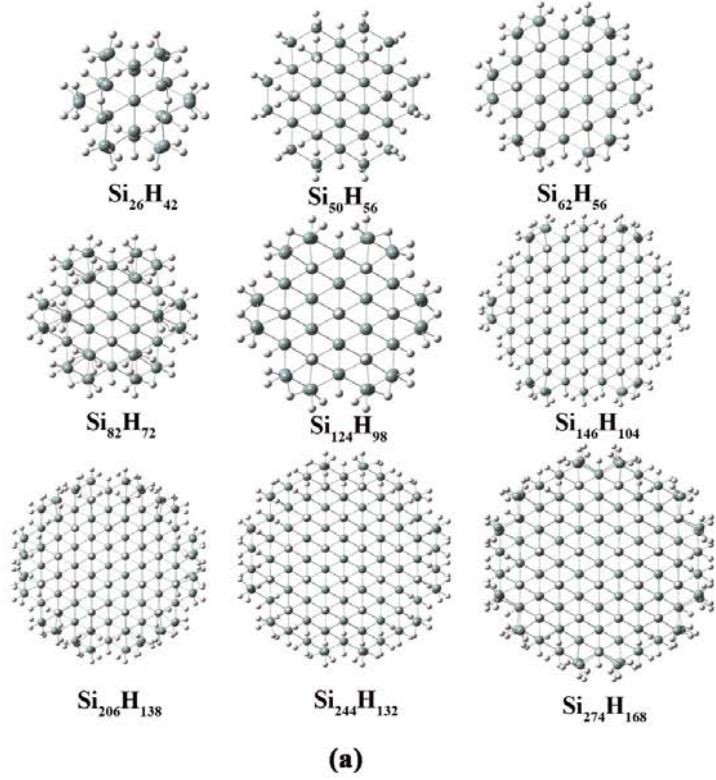

**(a)**

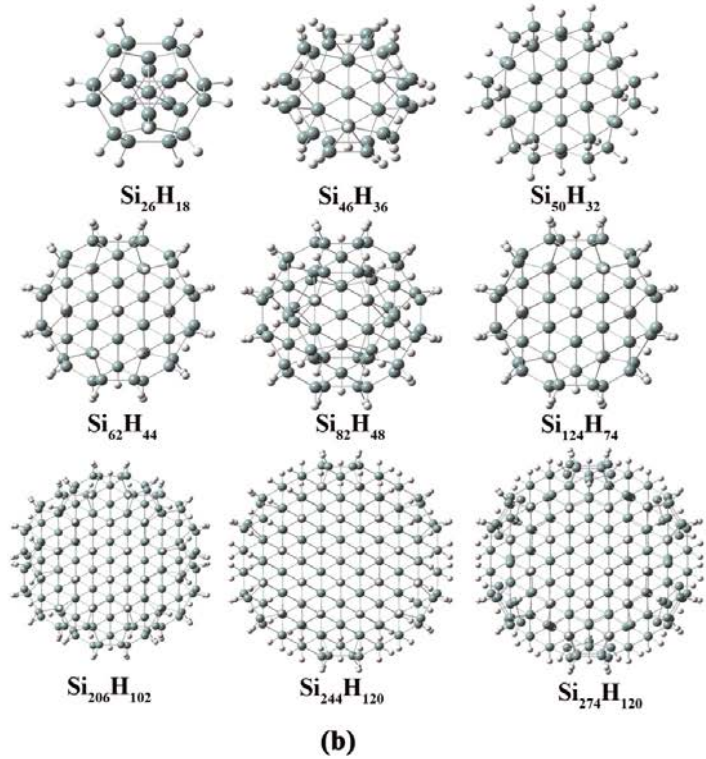

**(b)**

Figure 2: Optimized geometries of $Si_{26}H_{42}$, $Si_{50}H_{56}$, $Si_{62}H_{56}$, $Si_{82}H_{72}$, $Si_{124}H_{98}$, $Si_{146}H_{104}$, $Si_{206}H_{138}$, $Si_{244}H_{132}$ and $Si_{274}H_{168}$ elongated quantum dots (a) and $Si_{26}H_{18}$, $Si_{46}H_{36}$, $Si_{50}H_{32}$, $Si_{62}H_{44}$, $Si_{82}H_{48}$, $Si_{124}H_{74}$, $Si_{206}H_{102}$, $Si_{244}H_{120}$ and $Si_{274}H_{120}$ reconstructed quantum dots (b).

However, elongated quantum dots are grown along [111] by keeping two atoms in origin (instead of one). Elongated quantum dots range from 26 to 274 Si atoms including 42 to 168 H atoms (a total of 442 atoms). The diameter of the small ($Si_{26}H_{42}$) and large ($Si_{274}H_{168}$) dot is 10.05 Å and 20.98 Å, respectively. Furthermore, reconstructed quantum dots are also grown along [111] (with same technique we used for elongated dots) along with further surface reconstruction.[18] These dots range from 26 to 274 Si atoms, with 18 to 120 H atoms (a total of 394 atoms in largest dot). The diameter of the small ($Si_{26}H_{18}$) and large ($Si_{274}H_{120}$) dot is 10.66 Å and 20.06 Å respectively. All reconstructed dots are of $D_{3d}$ symmetry. For every distinct model, geometries have been fully optimized within symmetry constraints, using the hybrid B3LYP functional. The optimized structures of the elongated and reconstructed quantum dots are shown in Figure 2.

### 3.2. Electronic Properties

We investigate electronic properties for all stable spherical, elongated and reconstructed silicon quantum dots. Table 1 presents summary of all properties discussed in this study. Figure 3 shows diameter-dependent HOMO-LUMO gap of spherical quantum dots. The black dots represent our previous work and red curve shows QC fit (equation 1) whereas blue triangles represent our current work on large dots up to 32 Å (3.2 nm) in diameter. Concerning this plot, it is worthwhile to understand that large dots (blue triangles) are not included during quantum confinement fit (fit is only applied to the small dots).

Surprisingly, large dots follow the fit very well, which is clear evidence of an accurate formula. Based on our previous work[5] using quantum confinement concept, the "extrapolation formula" of our *ab initio* results can be described by the expected dependence of the HOMO-LUMO (also optical) gap on size (number of atoms or "diameter") as:

$$E(N) = A + B * N^{-n}$$

or                                                                                              (1)

$$E(D) = C + F * D^{-m}$$

where $A,\ B,\ n\ ,\ C,\ F$ and $m$ respectively are used as adjustable parameters to be determined by the fit. The $D$ is the diameter and/or $N$ is total number of silicon atoms of the quantum dot. Initially $m$ (and $n$) were free fitting parameters to be determined and the value obtained by the fit for $m$ was $m = 0.89 \pm 0.15$, whereas the values quoted in the literature vary between 0.76 and 1.3. The value obtained for the parameter $C$ by the same fit, which did not include the large dots, was $C = 1.02 \pm 0.25$ eV. As was explained in Ref. 5, this value of $C$ corresponds to the energy gap (band gap) of the infinite crystal, since as $D \rightarrow \infty$, $E$ becomes equal to $E(\infty)$, which, surprisingly enough, is in very good agreement with experiment. However, after inclusion of some (not all) of the larger dots in the fit, the quality of the fit ($\chi^2$) was improved and the value of the exponent $m$ was shifted towards 1 ($0.98 \pm 0.10$ ), which is highly suggestive that this exponent might have some kind of "universal" value equal to unity. One could rationalize this by considering the analogy between HOMO-LUMO gap (which is a measure of chemical hardness, or kinetic stability as shown in the article of Zdetsis,[34] and stability (cohesive stability) which is quantified by the cohesive energy i.e. the larger the gap, the larger the stability). In a recent paper by Zdetsis *et al.*[36] it was illustrated that cohesive energy of a nanocrystal varies inversely proportional to its diameter. Therefore, it is reasonable to assume that the HOMO-LUMO (energy) gap would also vary inversely proportional with the Diameter of the nanocrystal. Hence, in subsequent fits we have fixed the value of the exponent $m = 1$. With the same reasoning the value of the exponent $n$ was fixed to the value $n = 1/3$ (since the total number of silicon atoms, $N$, is proportional to the $3^{\text{rd}}$ power of the diameter $D$). As will be explained further below, the $N$-dependence of the gap can be described more accurately, compared to the $D$-variation, due to the uncertainties in defining the equivalent "diameter of the nanocrystal. Thus, the new $D$-dependence of the HOMO-LUMO gap for spherical dots in Figure 3 has the form:

$$E(D)_{HL,spherical} = (1.33 \pm 0.1) + (41.8 \pm 1.6) \times D^{-1} \tag{2}$$

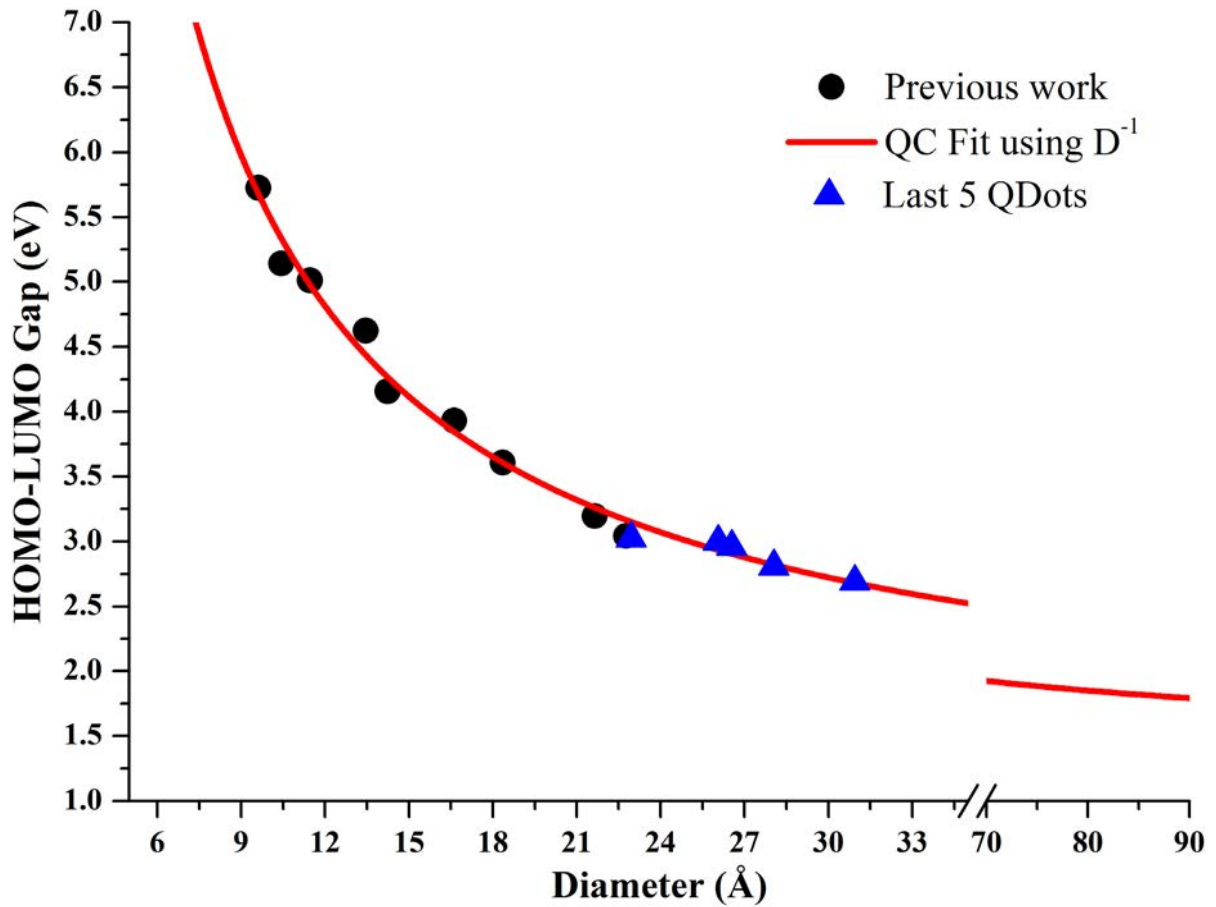

Figure 3: The plot shows HOMO-LUMO gap dependence on diameter of spherical quantum dots. The black dots represent our previous work[5], whereas blue triangles are from our current work (last five dots), in order to verify QC fit accuracy.

This new fit which includes in the fitting process the larger (but not the largest) dots is shown in Figure 4 together with an analogous fit for the elongated dots grown along the [111] direction. As would be expected for large diameters the two fits practically coincide and the trends, as well as the fitted parameters are the same within the (statistical) error margins, as we can see in relation 3. Small differences exist for small diameters due to small differences in the geometric arrangement and the "neighbourhood" around each silicon atom. Obviously, for very large dots these differences become marginal (and eventually zero).

$$E(D)_{HL,spherical} = (1.32 \pm 0.1) + (41.8 \pm 1.2) \times D^{-1}$$

(3)

$$E(D)_{HL,elongated} = (1.27 \pm 0.1) + (40.2 \pm 1.0) \times D^{-1}$$

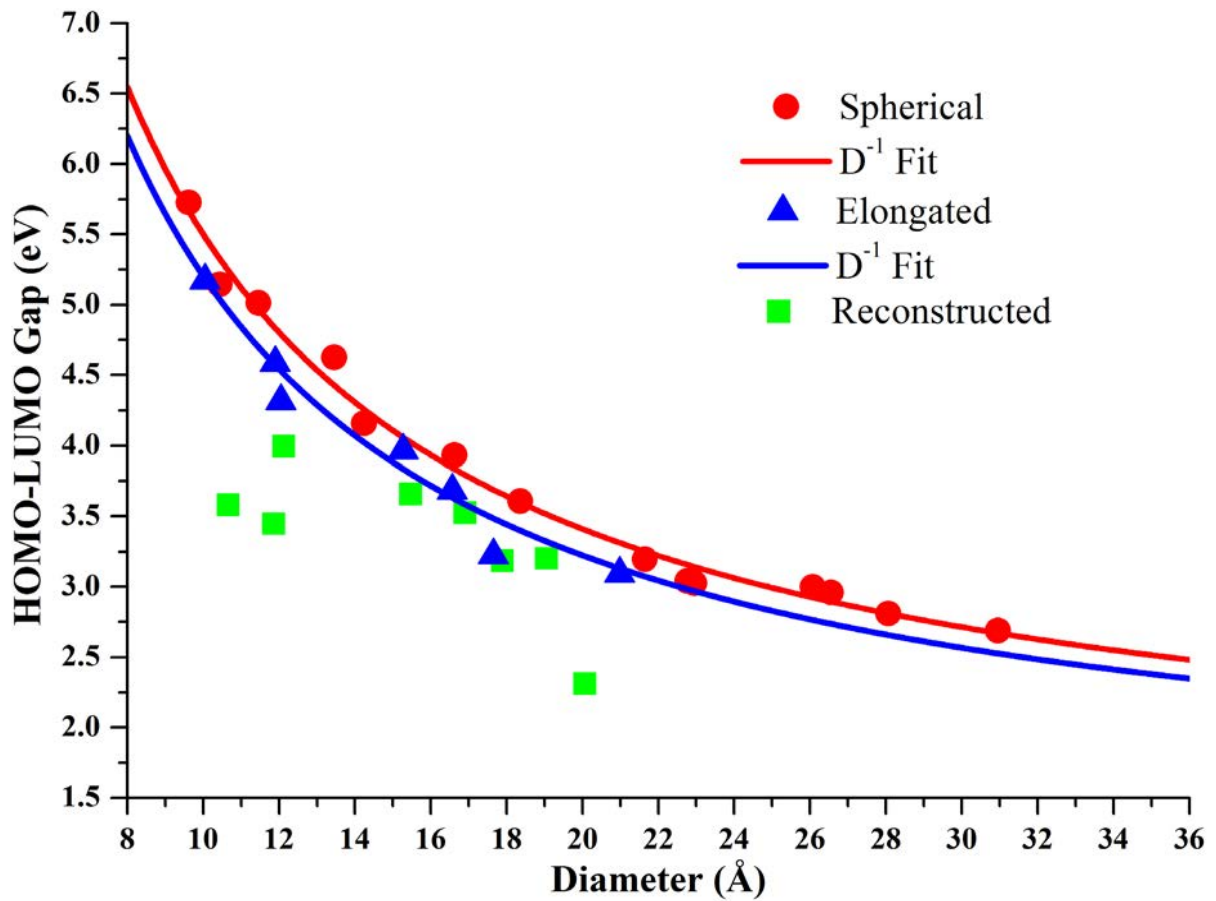

Figure 4: Representation of the HOMO-LUMO gap energy dependence on the diameter of the dots for spherical and elongated quantum dots.

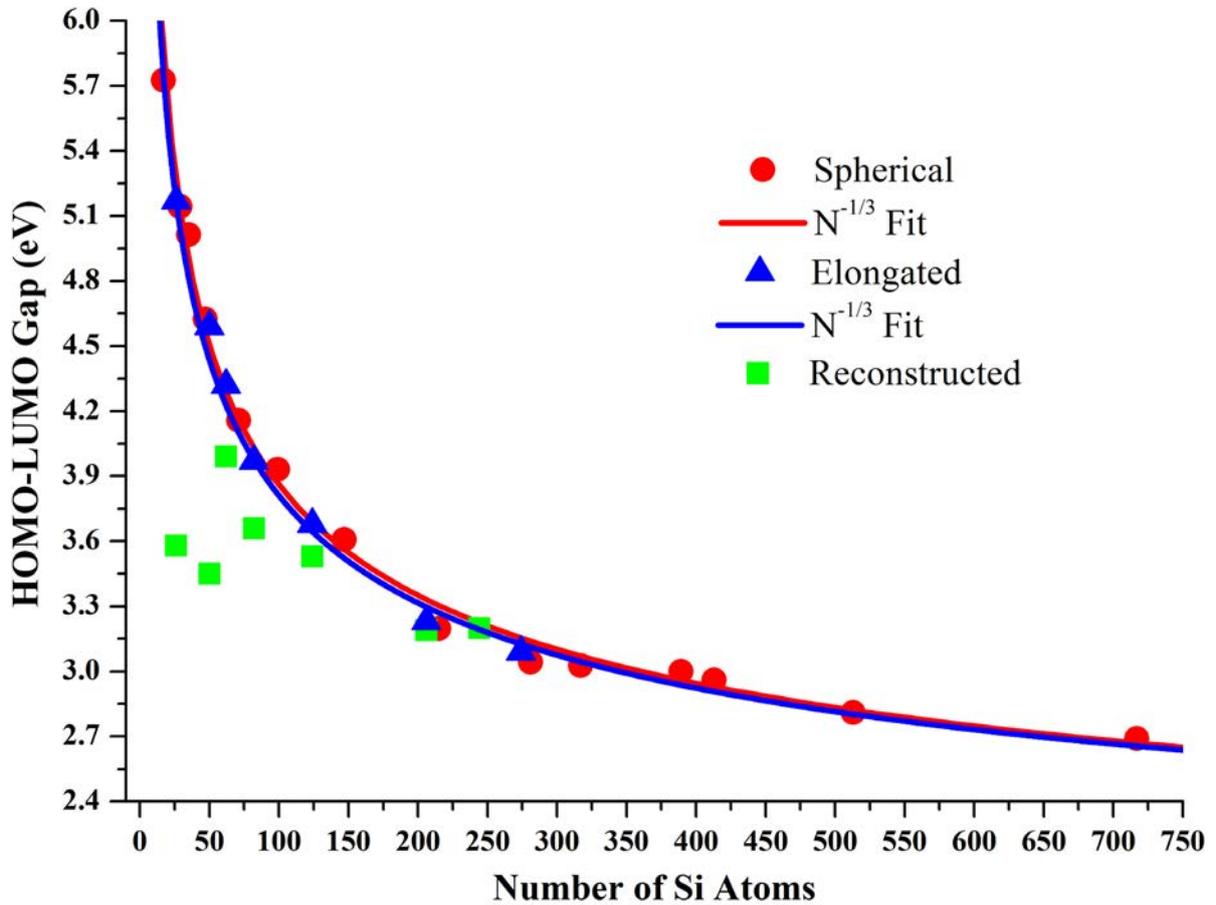

Figure 5: This plot corresponds to the HOMO-LUMO gap energy dependence on the number of Si atoms for spherical and elongated whereas reconstructed quantum dots are shown as random points.

     In Figure 4, together with the spherical and "elongated" dots we also display for comparison the HOMO-LUMO gaps of reconstructed dots as random points. As we can see in equation 3 the HL gap (parameter $C$) difference between spherical and "elongated" quantum dots is around 0.05 eV (gap difference of both infinite crystals). One can expect larger differences at small quantum dots which is obvious. For example, HOMO-LUMO gap difference between spherical and "elongated" dots at identical size of 8 nm is 0.09 eV (where

$C_{spherical} = 1.84 \, eV$ and $C_{elongated} = 1.75 \, eV$, respectively) which decreases with the increase in dots size and finally becomes zero (or nearly zero) at infinity. It is also worth to mention that the HOMO-LUMO gap value of infinite crystal 1.32 eV (or 1.27 eV for elongated) is larger than the experiment band gap value which is probably due to the negligence of many body effects in most of the DFT (electronic hence HL gap) calculations. For this reason we also carried out TDDFT calculations as well so that we can accurately compare experiment energy gap values with optical gap (section 3.3).

We have already explained above that the *N*-dependence of the gap can be described more accurately, compared to the *D*-variation therefore Figure 5 shows *N* - dependent energy gap (HOMO-LUMO gap) fit for spherical dots together with the elongated dots grown along the [111] direction and reconstructed dots. It is clear from the comparison between equations 3 and 4 that the energy gap values of infinite crystal (parameter *A*), for both spherical and elongated dots, are expectedly the same (within the statistical error margin).

$$E(N)_{HL,sherical} = (1.38 \pm 0.1) + (11.4 \pm 0.2) \times N^{-1/3}$$

$$E(N)_{HL,elongated} = (1.41 \pm 0.1) + (11.2 \pm 0.3) \times N^{-1/3}$$

(4)

Figure 6 shows distribution of highest occupied molecular orbitals (HOMOs) and lowest unoccupied molecular orbitals (LUMOs) for all three candidate quantum dots. As we can see in the figure, the HOMOs and LUMOs are mainly localized the interior of the dots for spherical and elongated quantum dots (without reconstructions). Whereas, after reconstruction, the HOMOs are localized inside and LUMOs distributed on the surface near the reconstruction sites of the quantum dots. This feature is also present in the work of Ref.

18 with which we agree. However, one must be careful when making such comparisons because the isovalue used for the drawing is very important. We can see in Figure 6 the representation drawn at <0.02 is different from expectations (as we observed in smaller dots <15 Å). Hence special care must be taken in making such graphical representations.

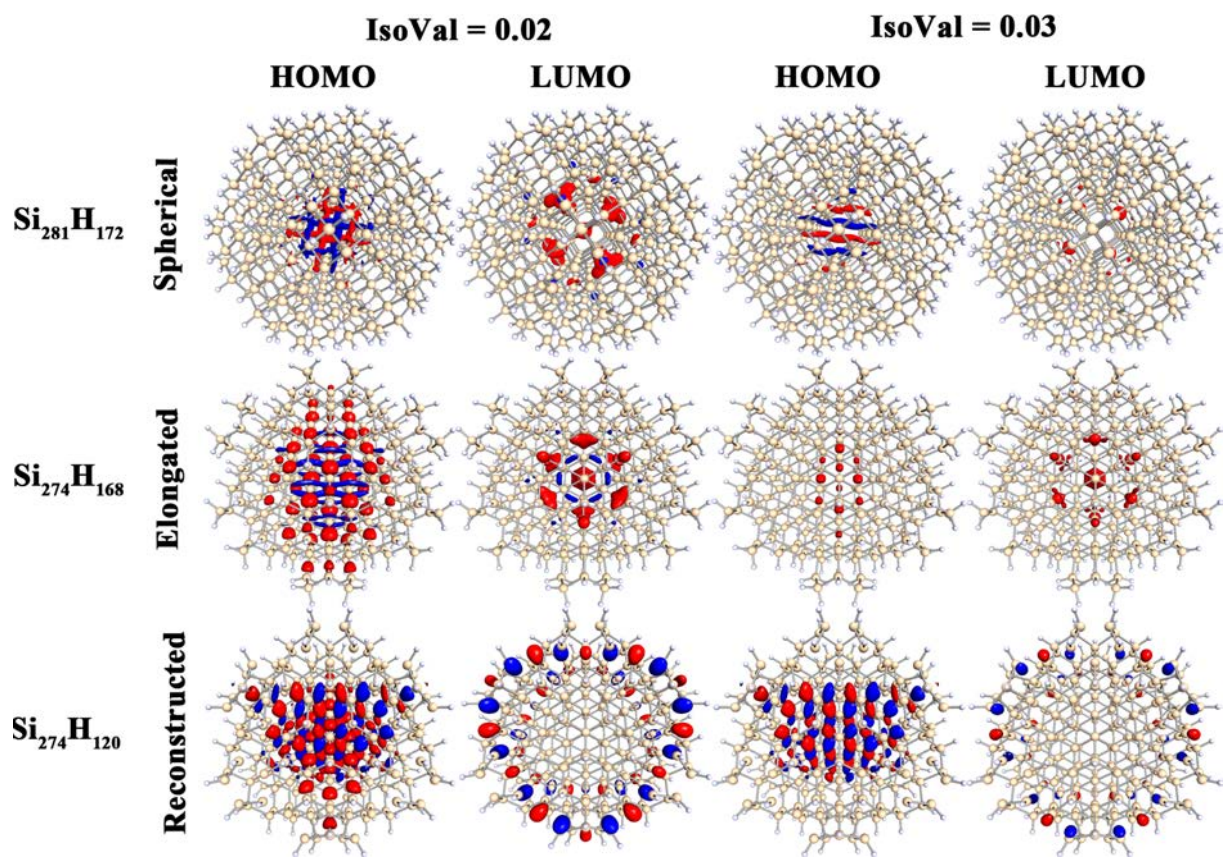

Figure 6: The highest occupied molecular orbital (HOMO) – lowest unoccupied molecular orbital (LUMO) graphical representation of $Si_{281}H_{172}$ (spherical), $Si_{274}H_{168}$ (Elongated) and $Si_{274}H_{120}$ (Reconstructed) quantum dots at iso-value 0.02 and 0.03 respectively.

Clearly, for small-medium size dots the spherical dots have larger gaps compared to reconstructed dots, and therefore on the basis of "kinetic stability" (or chemical hardness) would be expected to be more stable with highest HOMO-LUMO gap for a given diameter compared to elongated dots. For larger dots as $n \to \infty$ the results for spherical and elongated dots, as would be expected are practically identical. Concerning spherical dots, for which we have considered a much larger number of sizes (larger number of points in the graph), the results obtained here agree with our previous calculations[4,5] and with experimental results. For spherical and elongated dots quantum confinement fit produces excellent results with good match to the experimental values for HOMO-LUMO gap. Due to the good quality of the fit one can predict HOMO-LUMO gap for infinite system obviously, reconstructed dots are not expected to, and they do not follow such fitting scheme because of their random size dependence behavior for large sizes.

### 3.3. Optical Properties

We perform TDDFT/B3LYP/SVP level calculations for optical properties taking into account for spherical dots and some selected candidate dots of elongated and reconstructed growth models. We present optical properties in the form of diagrams of $N$-dependence and $D$-dependence of the optical gap (where $D$ and $N$ correspond to the diameter and number of heavy atoms of dots, respectively). In Figure 7, during the fitting process we consider our small spherical dots (black spheres)[5] and then we place larger dots (blue triangles) which nicely follow the fitting curve.

The infinite crystal optical gap compared with experiment results is in a very good agreement (larger dots are not included yet) hence results of fitting function can be observed:

$$E(D)_{OPT,spherical} = (1.14 \pm 0.1) + (37.3 \pm 0.4) \times D^{-1} \qquad (5)$$

**Table 1: Structural (total number of atoms, symmetry, diameter), Energetic (cohesive energy per silicon atom, binding energy per silicon atom), Electronic (HOMO, LUMO, H-L gaps from DFT calculations) and Optical (optical gap for TDDFT calculations) characteristics of spherical, reconstructed and elongated silicon quantum dots, respectively.**

| Qdots | Sym | Total # of atoms | Diameter (Å) | Cohesive Energy (eV/Si) | Binding Energy (eV/Si) | Homo (eV) | Lumo (eV) | H-L Gap (eV) | Optical Gap (eV) |
|---|---|---|---|---|---|---|---|---|---|
| $Si_{17}H_{36}$ | $T_d$ | 53 | 9.62 | 1.86 | 9.21 | -7.29 | -1.56 | 5.72 | 5.03 |
| *$Si_{29}H_{36}$ | $T_d$ | 65 | 10.44 | 2.78 | 7.09 | -6.84 | -1.69 | 5.14 | 4.52 |
| $Si_{35}H_{36}$ | $T_d$ | 71 | 11.46 | 3.03 | 6.60 | -6.70 | -1.69 | 5.01 | 4.39 |
| $Si_{47}H_{60}$ | $T_d$ | 107 | 13.45 | 2.77 | 7.20 | -6.55 | -1.93 | 4.62 | 4.02 |
| $Si_{71}H_{84}$ | $T_d$ | 155 | 14.24 | 2.84 | 6.95 | -6.38 | -2.22 | 4.15 | 3.59 |
| $Si_{99}H_{100}$ | $T_d$ | 199 | 16.62 | 3.05 | 6.55 | -6.24 | -2.31 | 3.93 | 3.40 |
| $Si_{147}H_{100}$ | $T_d$ | 247 | 18.36 | 3.41 | 5.77 | -6.00 | -2.39 | 3.60 | 3.12 |
| $Si_{215}H_{148}$ | $T_d$ | 363 | 21.64 | 3.40 | 5.79 | -5.84 | -2.65 | 3.19 | *2.79* |
| $Si_{281}H_{172}$ | $T_d$ | 453 | 22.77 | 3.48 | 5.60 | -5.76 | -2.72 | 3.04 | 2.62 |
| $Si_{317}H_{172}$ | $T_d$ | 489 | 22.95 | 3.56 | 5.44 | -5.73 | -2.71 | 3.02 | *2.69* |
| $Si_{389}H_{196}$ | $T_d$ | 585 | 26.07 | 3.62 | 5.35 | -5.82 | -2.82 | 2.99 | *2.49* |
| $Si_{413}H_{196}$ | $T_d$ | 609 | 26.55 | 3.66 | 5.28 | -5.81 | -2.85 | 2.96 | *2.46* |
| $Si_{513}H_{252}$ | $T_d$ | 765 | 28.06 | 3.64 | 5.32 | -5.70 | -2.89 | 2.81 | *2.39* |
| $Si_{717}H_{300}$ | $T_d$ | 1017 | 30.95 | 3.72 | 5.15 | -5.69 | -3.00 | 2.69 | *2.26* |
| $^\$Si_{26}H_{18}$ | $D_{3d}$ | 44 | 10.66 | 3.13 | 5.54 | -5.82 | -2.24 | 3.58 | 2.96 |
| $Si_{50}H_{32}$ | $D_{3d}$ | 82 | 11.86 | 3.33 | 5.56 | -5.76 | -2.31 | 3.45 | - |
| $Si_{62}H_{44}$ | $D_{3d}$ | 106 | 12.13 | 3.32 | 5.79 | -6.17 | -2.18 | 3.99 | - |
| $Si_{82}H_{48}$ | $D_{3d}$ | 130 | 15.47 | 3.43 | 5.47 | -6.07 | -2.41 | 3.66 | - |
| $Si_{124}H_{74}$ | $D_{3d}$ | 198 | 16.90 | 3.44 | 5.52 | -5.97 | -2.44 | 3.53 | - |
| $Si_{206}H_{102}$ | $D_{3d}$ | 308 | 17.89 | 3.56 | 5.29 | -5.65 | -2.46 | 3.19 | 2.79 |
| $Si_{244}H_{120}$ | $D_{3d}$ | 364 | 19.05 | 3.61 | 5.31 | -5.78 | -2.58 | 3.20 | - |
| $Si_{274}H_{120}$ | $D_{3d}$ | 394 | 20.06 | 3.59 | 5.12 | -5.58 | -3.27 | 2.31 | - |
| $^\#Si_{26}H_{42}$ | $D_{3d}$ | 68 | 10.05 | 2.39 | 8.01 | -6.80 | -1.63 | 5.17 | 4.47 |
| $Si_{50}H_{56}$ | $D_{3d}$ | 106 | 11.9 | 2.92 | 6.82 | -6.47 | -1.88 | 4.59 | - |
| $Si_{62}H_{56}$ | $D_{3d}$ | 118 | 12.05 | 3.16 | 6.30 | -6.37 | -2.05 | 4.32 | - |
| $Si_{82}H_{72}$ | $D_{3d}$ | 154 | 15.27 | 3.19 | 6.24 | -6.19 | -2.22 | 3.97 | - |
| $Si_{124}H_{98}$ | $D_{3d}$ | 222 | 16.57 | 3.28 | 6.03 | -5.99 | -2.31 | 3.68 | - |
| $Si_{206}H_{138}$ | $D_{3d}$ | 344 | 17.65 | 3.41 | 5.74 | -5.84 | -2.61 | 3.23 | 2.84 |
| $Si_{274}H_{168}$ | $D_{3d}$ | 442 | 20.98 | 3.48 | 5.61 | -5.77 | -2.68 | 3.09 | - |

*Stokes Shift = 2.07 eV          $Stokes Shift = 0.38 eV          #Stokes Shift = 2.09 eV

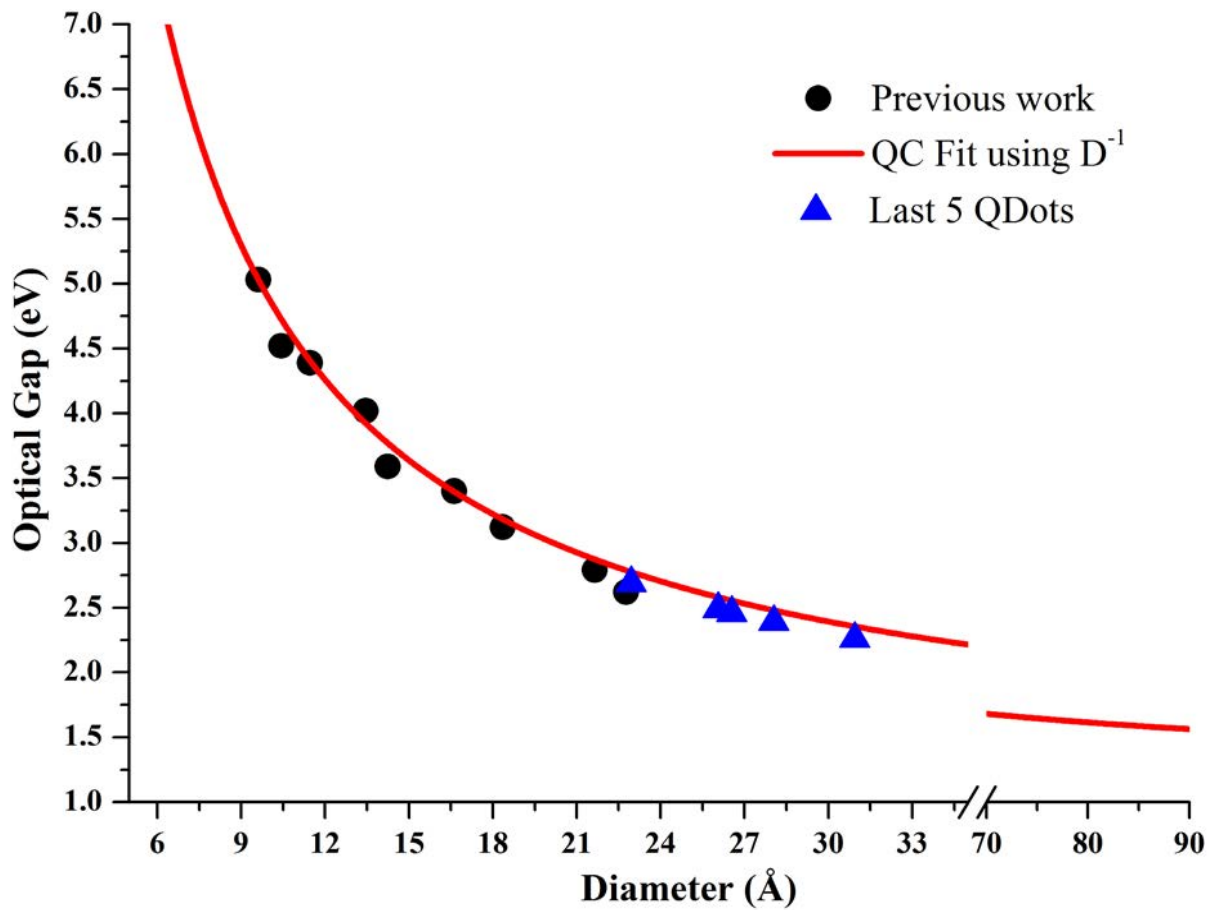

Figure 7: The plot shows optical gap versus diameter of spherical quantum dots. The black dots represent our previous work[5], whereas blue triangles are from our current work (last five dots).

The results in Figure 8 include in the fitting process large spherical dots (in addition to the small ones which are shown in Figure 7), which were not included in the fitting process of Figure 7. In the same we include some experiment results from different research groups for comparison (Figure 8). First of all, as we can see in the figure, the dispersion of the experimental data for dots of similar or equal sizes is impressive indeed. Besides uncertainties in the determination of diameters, such scattering of the experimental data is related also to different ways of preparation (for example gas phase, solution, etc.) and various surface conditions (ligands, etc). Our data and their expansion, through their fitting scheme, completely free from any influence or bias from experimental data, constitute true reference points and testing grounds, serving also as clear and accurate guiding lines and virtual "yard sticks" from all sizes. This is because the fit and the resulting interpolation-extrapolation formula are of excellent quality, as can be seen also from the numerical values (and interpretation) of the fitted parameters and their statistical error bars:

$$E(D)_{OPT,spherical} = (1.08 \pm 0.1) + (36.3 \pm 0.5) \times D^{-1} \qquad (6)$$

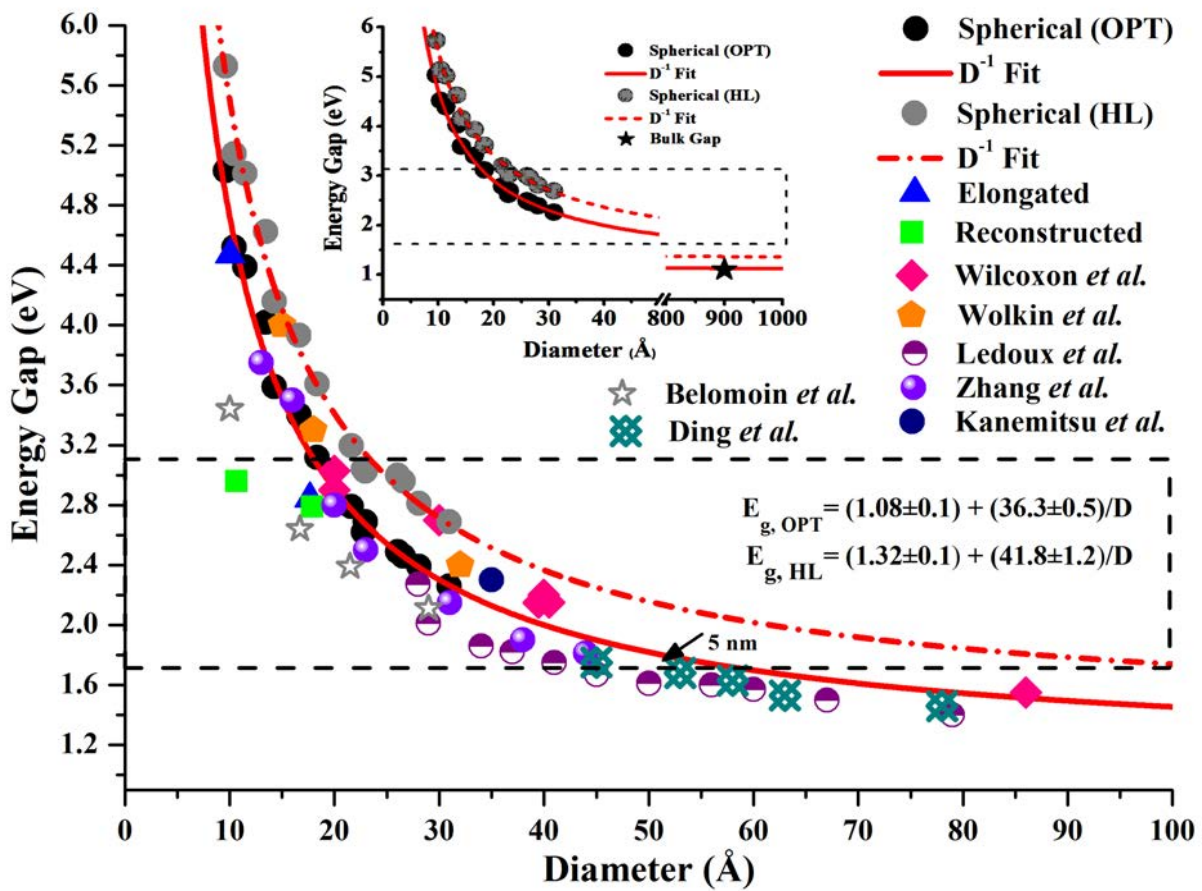

Figure 8: The plot shows energy (optical and H-L) gap with respect to the diameter of the spherical quantum dots (red) along with few point from reconstructed (green) and elongated (blue) quantum dots. We also include some experiment results[19-25] from different research groups for comparison. Dotted rectangle correspond to visible light range and black star (inset) shows experiment band gap value for silicon.

As can be seen, we have obtained the experimental value of the gap with almost "chemical accuracy". Clearly, the optical (energy) gap value, which is 1.08 eV, is improved compared to the previous fitting results (Figure 7) and in a perfect agreement with the experiment value (i.e 1.1 eV). As we can see in the figure, the randomly chosen small elongated dot shows slightly smaller optical gap for that particular size, compared to the corresponding spherical dot. For larger sizes, as $D \rightarrow \infty$, the results will coincide.

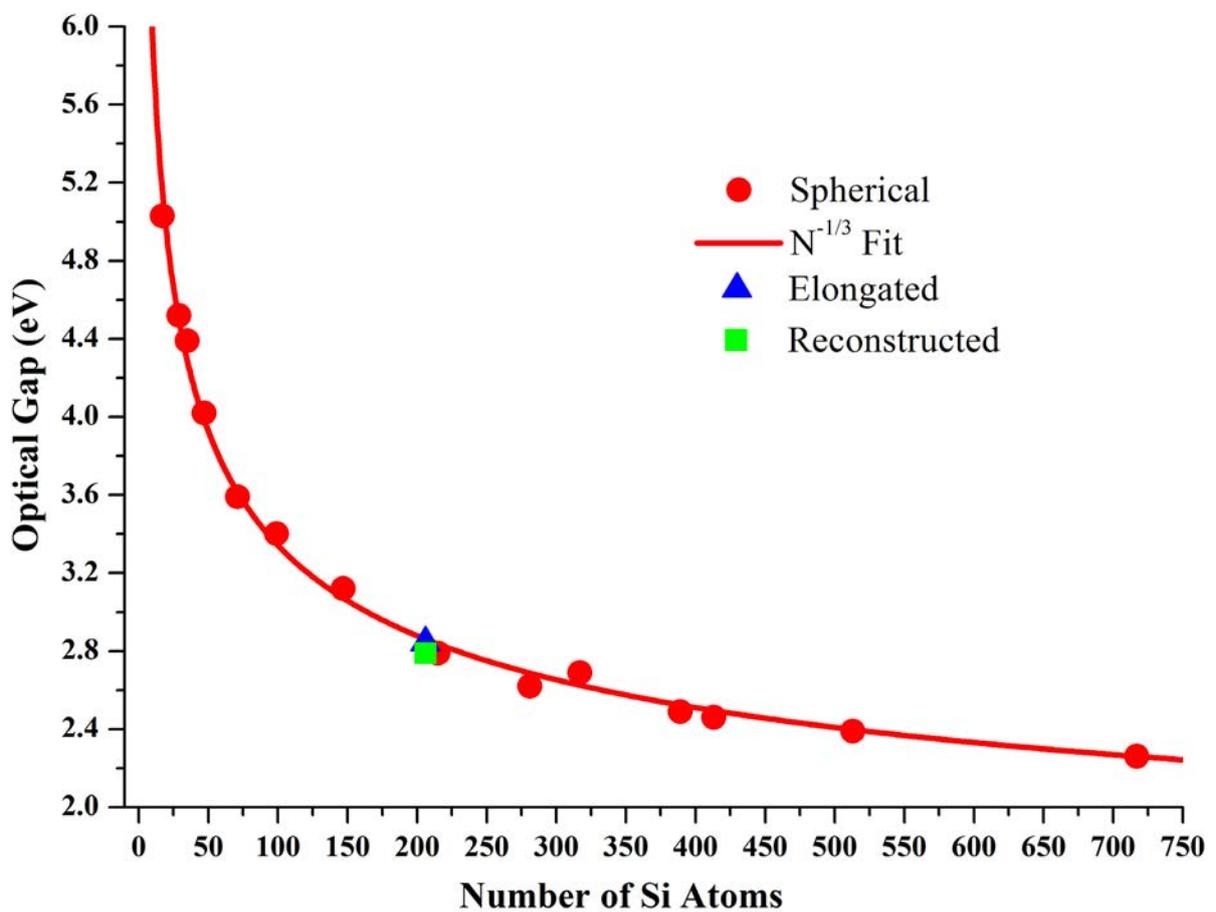

Figure 9: This plot shows optical gap energy dependence on the number of Si atoms for spherical quantum dots whereas one point from reconstructed and elongated quantum dots represents comparison.

In addition to the *D*-dependence of the optical gap, Figure 9 shows the variation of the optical gap in terms of the number of silicon atoms of spherical quantum dots. The quality of the fit, as shown above (for the D-dependence) is excellent and even more improved, since there is not additional uncertainty about the exact determination of the diameter D. The new parameters are given by the following equation:

$$E(N)_{OPT,sherical} = (1.09 \pm 0.1) + (10.4 \pm 0.1) \times N^{-1/3} \qquad (7)$$

The agreement with experiment is excellent.

We also present IR spectra in Figure 10 of one candidate structure of each growth model. First peaks correspond to the Si-H bonding frequencies and second large peaks shows Si-Si bond frequencies.

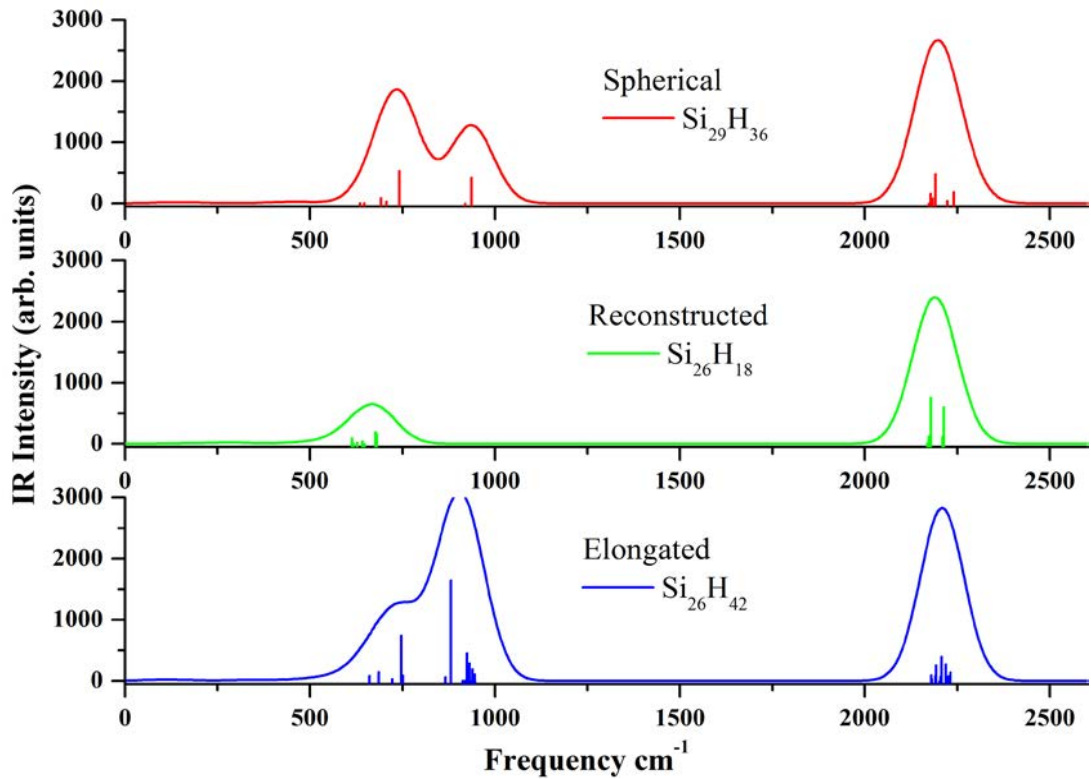

Figure 10: IR spectrum of spherical Si$_{29}$H$_{36}$, reconstructed Si$_{26}$H$_{18}$ and elongated Si$_{26}$H$_{42}$ quantum dots. The continuous curves are produced by Gaussian broadening. Calculations were performed at B3LYP/SVP level.

### 3.4. Cohesive Properties

In this section we discuss cohesive/binding/atomization properties of our spherical, elongated and reconstructed quantum dots. We investigate cohesive properties by calculating binding/atomisation energy and cohesive energy per silicon atom as a function of number of silicon atom (or $N$-dependence of binding and cohesive energy using equation 1) for all three growth models (see Figures 11, 12). Binding energy of a quantum dot is defined as:

$$BE_{QD} = N_{Si}E(Si) + N_H E(H) - E_{QD}[Si_{N_{Si}}H_{N_H}]$$ (8)

where $N_{Si}$ and $N_H$ represents total number of silicon and hydrogen atoms respectively within silicon quantum dot, $E(Si)$ and $E(H)$ are energies of single silicon and hydrogen atom respectively and $E_{QD}[Si_{N_{Si}}H_{N_H}]$ represents total energy of the quantum dot. Figure 11 shows binding energy per silicon atom as function of the size of spherical, elongated and reconstructed quantum dots. In this diagram, the reconstructed dots are artificially appearing as less stable compared to spherical and elongated dots, because the contribution of surface hydrogens is not taken fully and correctly into account. When this is properly done (with the definition of cohesive energy), as will be shown below, the reconstructed dots, would be clearly more stable as would be expected. In agreement with the relative size of the HOMO-LUMO gap, we can verify in Figure 11 that the spherical dots (for small and medium sizes) are more stable compared to the elongated dots. Furthermore, as was explained earlier in the discussion for the size variation of the energy gap, the binding energy should follow a $N^{-1/3}$ size dependence.[35,36] As was demonstrated by Zdetsis *et al.*,[35,36] this type of fitting process can reproduce with chemical accuracy the cohesive energy of the infinite crystal, provided the proper functional (or meta-functional) has been chosen. The parameters obtained from the fit for the binding energy using the B3LYP functional are:

$$E(N)_{B,sherical} = (4.13 \pm 0.1) + (9.9 \pm 0.6) \times N^{-1/3}$$

$$(9)$$

$$E(N)_{B,elongated} = (4.14 \pm 0.3) + (9.2 \pm 1.0) \times N^{-1/3}$$

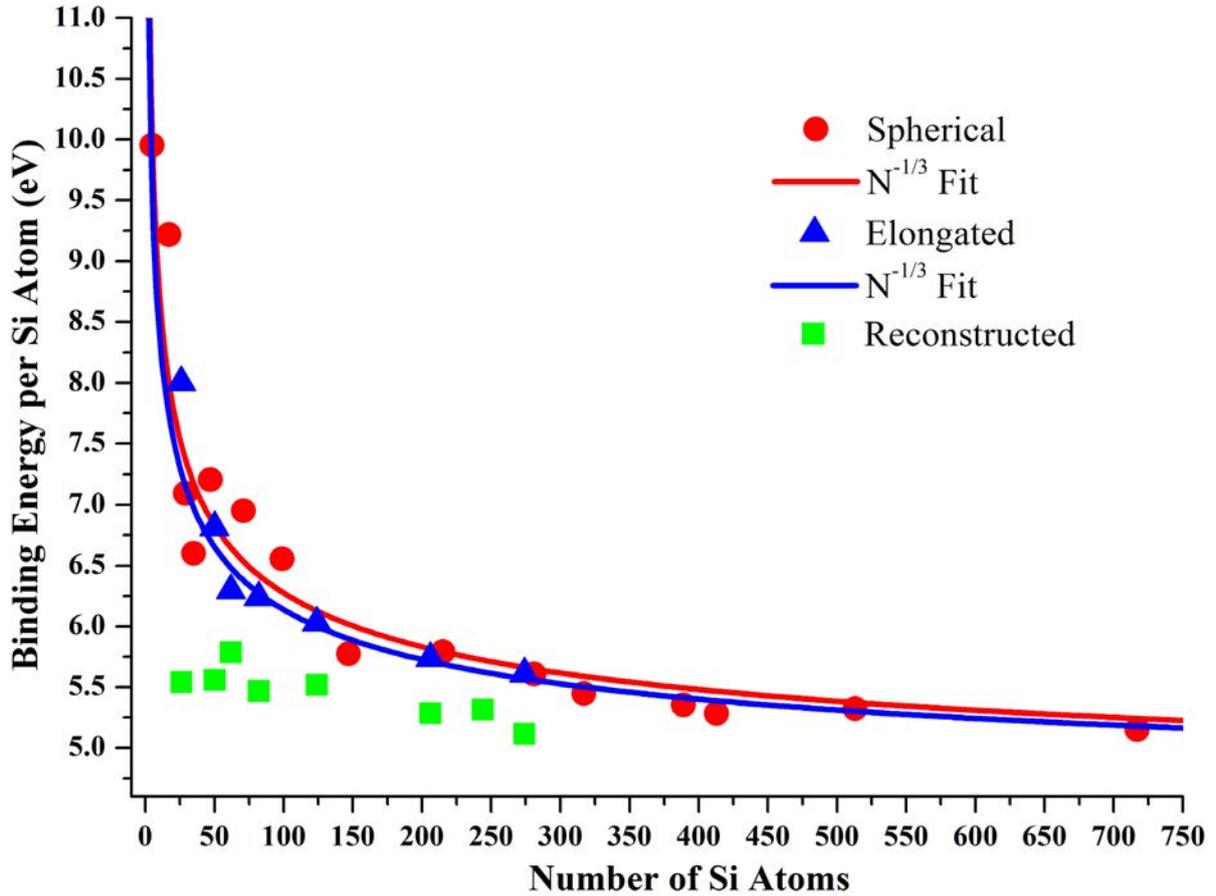

Figure 11: This plot presents binding energy per Si atom depending on number of Si atoms for spherical, elongated and reconstructed quantum dots.

The extrapolated values of the binding energy of the infinite crystal are 4.13 eV/atom and 4.14 eV/atom for spherical and elongated dots respectively, whereas the experimental value is 4.63 eV/atom.[37] This result is quite good (especially when considering the simplicity of the method, versus other high level direct methods) but is not so "spectacular" as for the energy gap, for which the superiority of the B3LYP functional for silicon has been demonstrated earlier by Zdetsis *et al.*[34] Apparently, B3LYP is not as good for cohesive properties as for electronic and optical properties. This is also verified by the plot (and fit) of the "cohesive energy" (instead of binding energy per atom), shown in Figure 12 below.

The cohesive energy deals with the binding energy of the Si core for which the interaction of the surface atoms have been taken into account in a uniform way through the introduction of the Chemical potential of hydrogen. This way, reconstructed nanocrystals, which by construction have a much smaller number of passivating hydrogen atoms, would be naturally appeared more stable, which is obviously true. Thus, although reconstructed nanocrystals, which technically have appeared to be less stable on the basis of binding energy per (Si) atom, would be more stable by the use of cohesive energy. As we can see in Figure 12 this is clearly so. The cohesive energy is defined as:

$$E_{Coh,QD} = [BE_{QD} + \mu_H N_H] \qquad (10)$$

where, $BE_{QD}$ is the binding energy of the quantum dot, $\mu_H$ is chemical potential of hydrogen and $N_H$ is total number of hydrogen atoms in quantum dot. Figure 12 shows the plot of cohesive energies versus the number of silicon atoms, fitted to the same $N^{-1/3}$ linear dependence as the binding energy per atom. The new parameters have been somewhat improved as can be seen in the following relations:

$$E(N)_{COH,sherical} = (4.21 \pm 0.1) + (-4.4 \pm 0.1) \times N^{-1/3}$$

$$\qquad (11)$$

$$E(N)_{COH,elongated} = (4.23 \pm 0.1) + (-4.7 \pm 0.5) \times N^{-1/3}$$

As we can see, these cohesive energy values for the infinite crystal are closer to the experimental values, compared to the binding energy per atom, but still not as satisfactory as the energy gap or the cohesive energies of the $MgH_2$ and $BeH_2$ crystals.[35,36] Nevertheless, it has been demonstrated here that the cohesive energy, for obvious reasons, is a better criterion of stability for Si nanocrystals (with a varying number of surface hydrogens) compared to binding energy per (Si) atom.

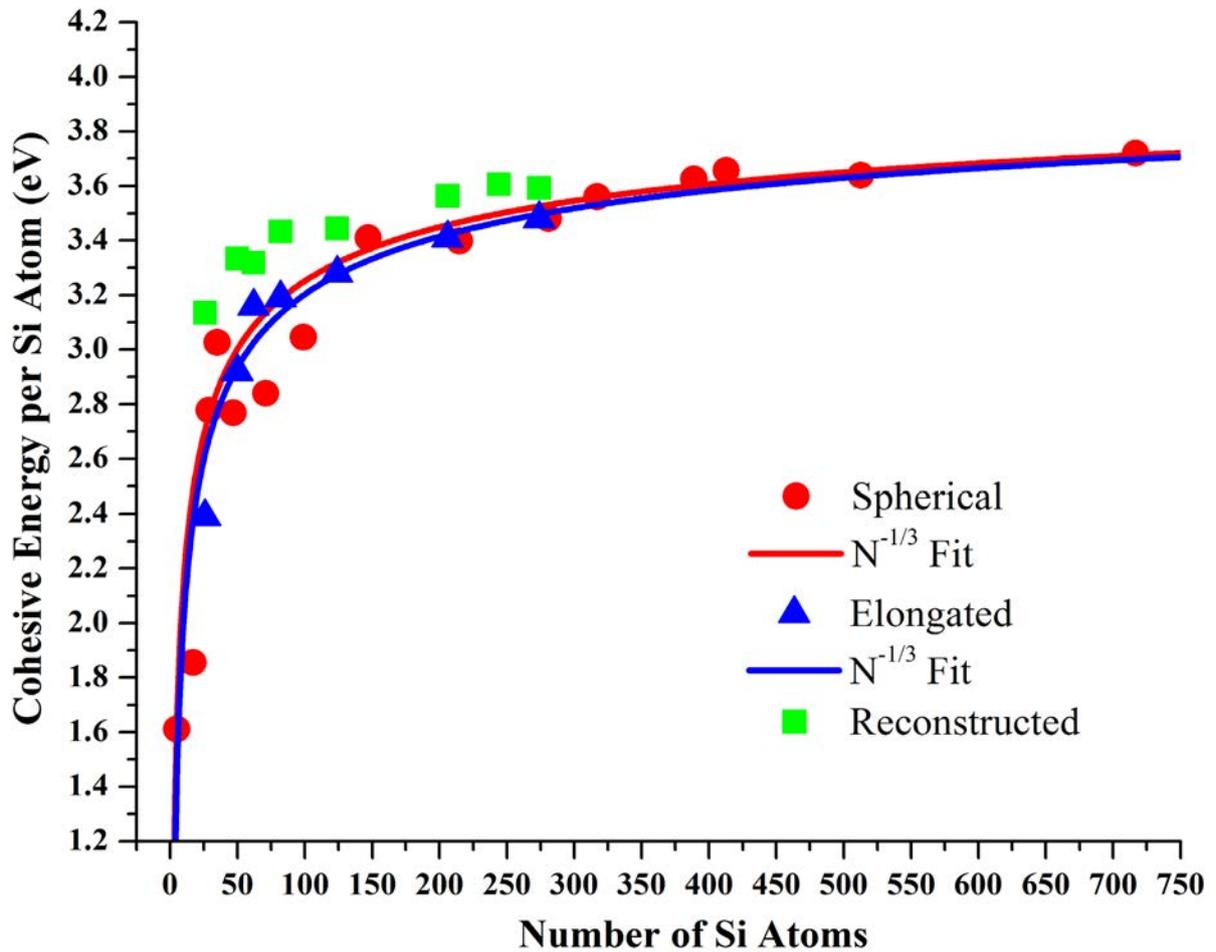

Figure 12: This plot corresponds to the cohesive energy per Si atom dependence on the number of Si atoms for spherical, elongated and reconstructed quantum dots.

To improve the calculated (extrapolated) cohesive energy of the infinite crystal we have, recalculated below the cohesive (and in part the optical) properties of selective silicon dots of small, medium and large (but not very large, due to computational cost) sizes using the M06 meta-functional, which was very successful for $MgH_2$ and $BeH_2$ nanocrystals.

## 3.5. Comparison with the M06 Meta-Functional[38]

For the reasons explained above, we have chosen to compare the results for silicon nanocrystals obtained with the B3LYP functional with similar results using the M06 (Meta) functional, using the same fitting scheme and procedures.

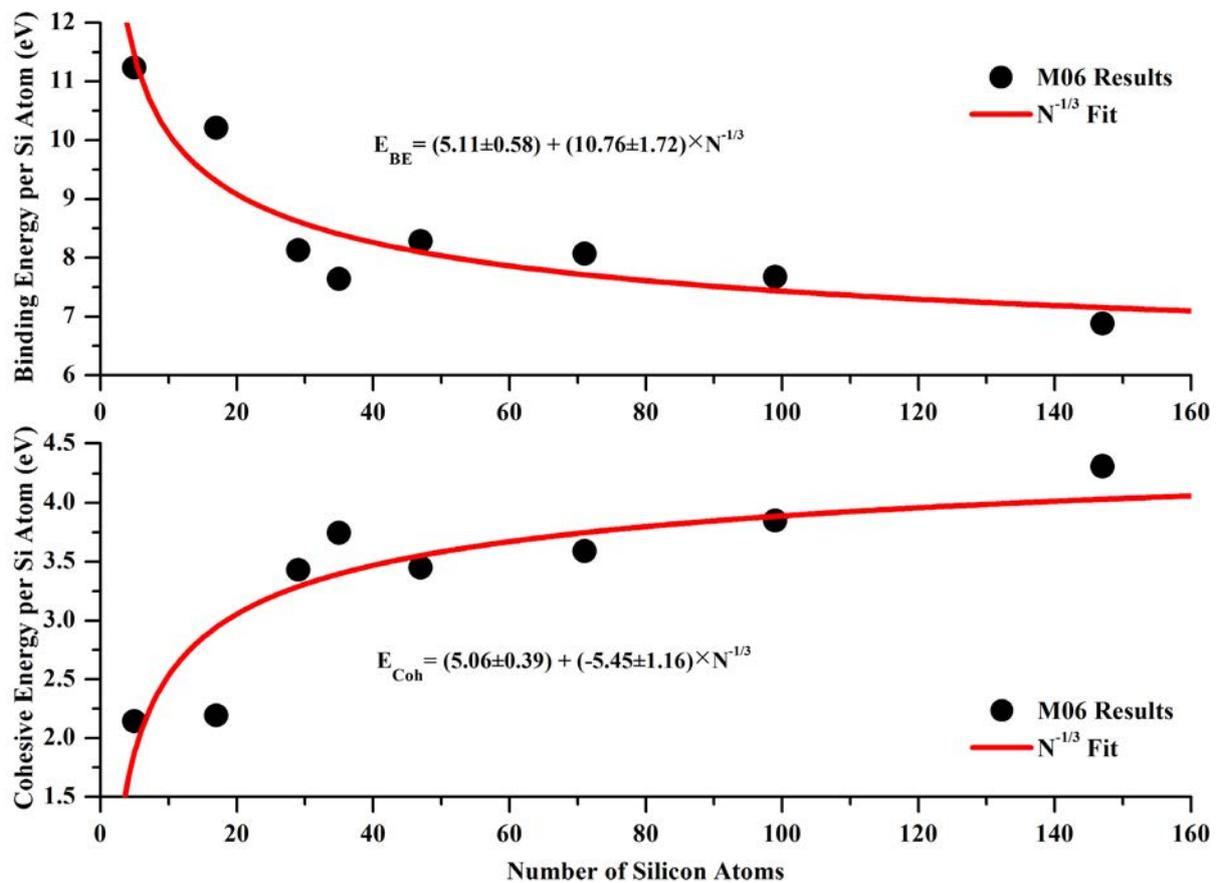

Figure 13: N-dependence of the binding energy per heavy atom and cohesive energy per heavy atom comparison using M06 functional.

Figure 13 represents the N-dependence of the cohesive and binding energy per silicon atom of small and medium size dots, using the M06 functional. As we can see, in contrast to Figures 11 and 12 (with B3LYP results), the M06 functional significantly improves (increases) both binding and cohesive energy per atom of the infinite system. However, at the same time it overestimates both of them. As was mentioned by Zdetsis *et al.*,[35,36] the M06 functional is very reliable for cohesive properties, but for silicon it seems that its success is not as dramatic as for the metal hydrides. This is partly due to inadequate number of points in the fit (as is indicated by the larger uncertainties in the infinite energy parameter $A$, $\pm 0.58$ and $\pm 0.39$ eV in both cases compared to $\pm 0.1$ and $\pm 0.1$ eV for B3LYP). Needless to say, that the experimental value of 4.63 eV for the cohesive energy is in fact obtained within the calculated statistical uncertainty of $\pm 0.4$ or $\pm 0.5$ eV, but this is not enough. Another possible reason for the not so good performance of the M06 functional is the geometry re-optimizations we have performed with this functional to obtained the (new) equilibrium geometries corresponding to the calculated cohesive and binding energies. As was shown by Zdetsis *et al.*,[35,36] although the M06 functional is very good for energies, is not so good for geometries. This is why in the calculations of cohesive properties in Refs. 35, 36, the geometries were optimized using the PBE functional, before the energies were computed by single point M06 calculations. It is anticipated that when recalculate the equilibrium geometries and add more points in the fit, the cohesive energy would improve significantly (perhaps close to the experimental value). This remains to be seen in future work.

In addition to the comparisons of cohesive and binding energies we have also performed comparisons for the electronic and optical gaps, for which we have already shown the excellent performance of the B3LYP functional. The results of such comparisons are shown in Figure 14, which shows both D-dependence and N-dependence of the HOMO-LUMO gap of small and medium size (limited data sets) spherical silicon quantum dots. Red squares represent B3LYP results and blue squares show M06 results whereas red and blue curves correspond to the fit using equation 1. Clearly, the M06 functional significantly overestimates the HOMO-LUMO gap of each of the dots, and consequently the infinite band gap compared with B3LYP functional. Therefore the comparison with experimental energy gap is getting worse. Again, avoiding the M06 geometry optimization could improve the results and the agreement with experiment, since M06 is not so good for geometry optimization (we can observe in Figure 14 that M06 slightly reduces the Si-Si bond length, as can be seen by the differences in the diameters of nanocrystals containing the same number of Si and H atoms).

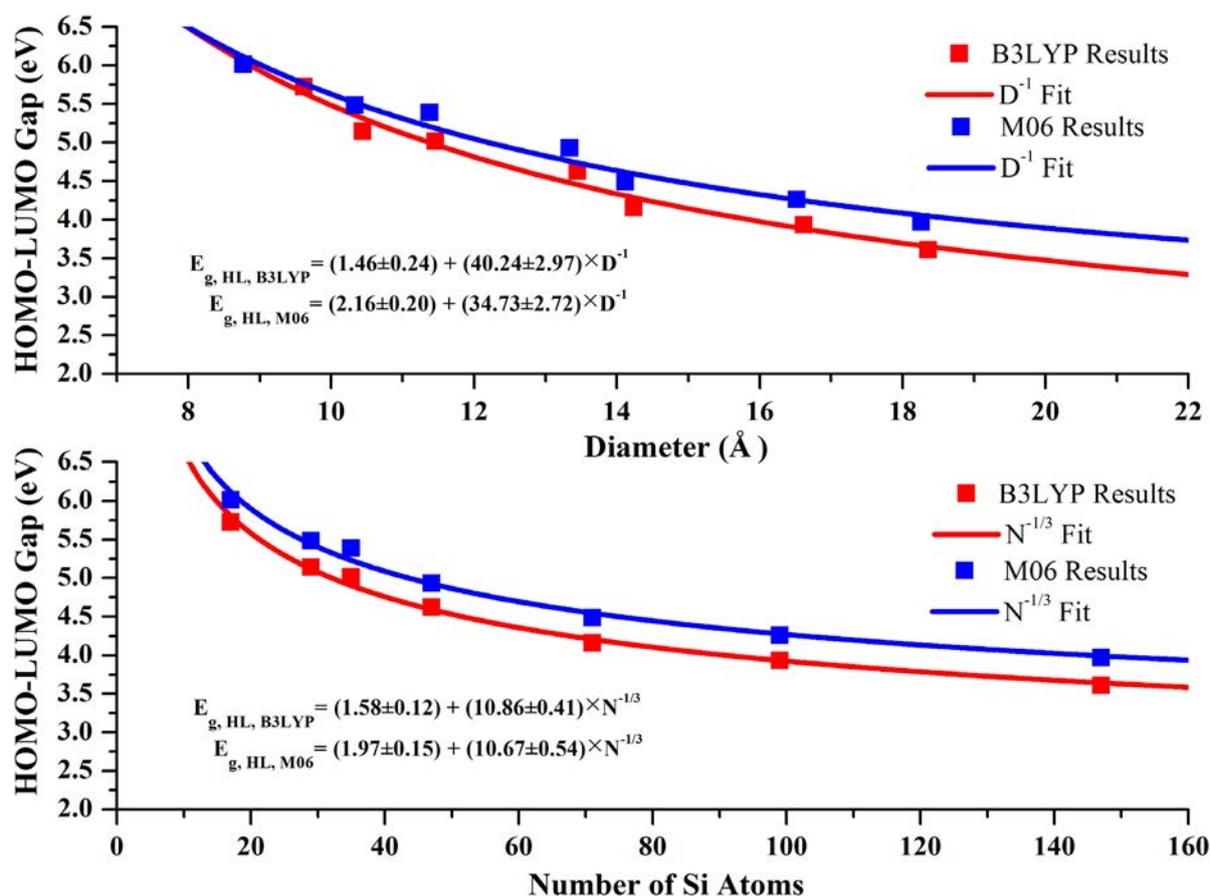

Figure 14: D-dependence and N-dependence of HOMO-LUMO gap comparison using B3LYP and M06 functionals.

## 3.6. Other Related Properties

There are in the literature several empirical (to one degree or another) relationships[26] connecting the dielectric constant (and /or the index of refraction) for a substance with the band gap as the key quantity. For example

$$n = \sqrt{1 + \left(\frac{A}{E_g + B}\right)^2} \qquad (12)$$

where $A = 13.6\ eV$ and $B = 3.4\ eV$ .

Therefore, from the above empirical relation (and its various extensions) one could have quick estimates of $n$ (and epsilon) if one wishes to rely in such methods. For obvious reasons, in our fully *ab initio* study here we have not attempted to obtain any such estimates.

## 4. CONCLUSIONS

In conclusion, we have thoroughly and systematically studied the structural, cohesive, electronic, and optical properties of small, medium and large silicon quantum dots, up to 32 Å in diameter (a total of 1017 atoms i.e. 717 silicon atoms and 300 hydrogen atoms) in terms of size, growth pattern and model description. Our results are fully consistent with the quantum confinement interpretation. An important, perhaps the most important, outcome of such study, besides the very satisfactory agreement with experimental measurements for nanocrystals (up to 32 Å in diameter), is the judicious extrapolation of the nanoscale results all the way to infinite silicon crystal, and the successful comparison with experiment for both the energy gap and the cohesive energy of crystalline silicon. We have found that the 1/3 expected dependence[36] of the cohesive energy on the number of particles can be fully appropriate and compatible with the gap size dependence on the grounds that "kinetic stability" and cohesive stability should vary in parallel, although this is not always valid.[34] The optical gaps of the nanocrystals, calculated with TDDFT, lead naturally (in an unbiased way) to the prediction of the band gap of crystalline silicon with almost chemical accuracy. The cohesive energy of the infinite crystal has been also obtained with very good accuracy, which, if needed, can be further improved in a systematic way. Our present results for the band gaps, which are based on our earlier findings for spherical Si quantum dots up to 20 Å in diameter, are in full agreement with those results and predictions. Thus our results can serve as a "yard stick" for quick (and rather accurate) estimate of such fundamental quantities.

Comparing the three different growth models (spherical, elongated and reconstructed) for small and medium size dots, the reconstructed nanocrystals are more stable in comparison to both unreconstructed ones. The spherical are more stable (and with larger HOMO-LUMO gaps) compared to the elongated dots. However, for large enough nanocrystals the stability and energy gaps become similar, and for very large dots (as $n \to \infty$) the results practically coincide, as would be expected.

**AUTHOR INFORMATION**

**Corresponding Author**

*E-mail: shanawersi@gmail.com. shanawersi@fen.bilkent.edu.tr. Phone: +90-312-290-2514. Fax: +90-312-266-4579.

**Notes**

The authors declare no competing financial interest.

**ACKNOWLEDGMENT**

The State Scholarships Foundation (Ιδρύμα Κρατικών Υποτροφιών) Greece "ΙΚΥ" and The Scientific and Technological Research Council of Turkey "TÜBİTAK" are gratefully acknowledged.

**Table of Contents Image**

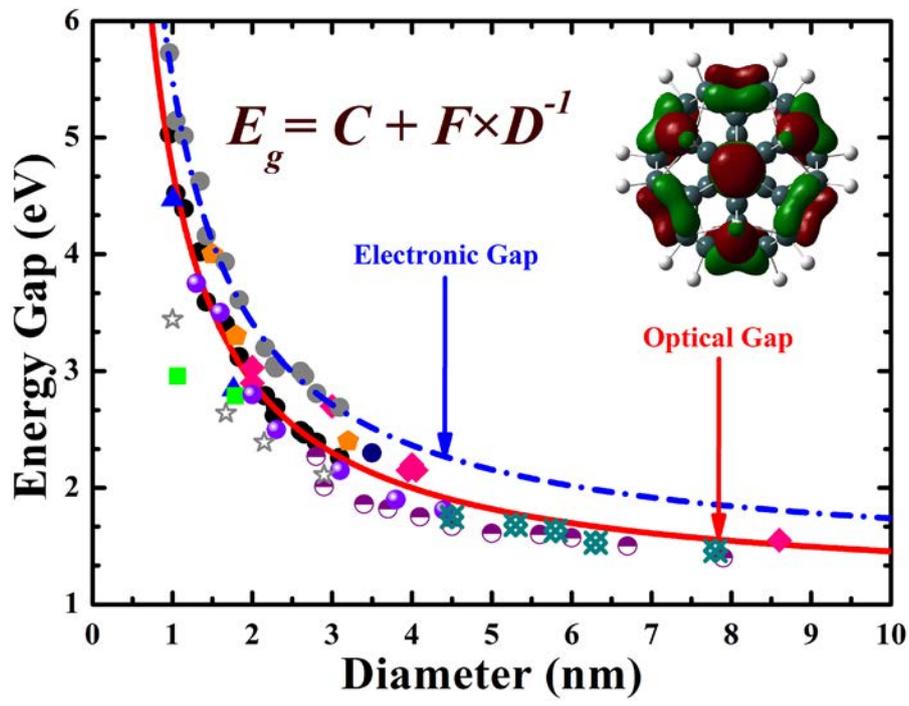